\documentclass[showpacs,floatfix,amsmath,amssymb, aps,twocolumn]{revtex4-1}

\usepackage{graphicx}% Include figure files
\usepackage{dcolumn}% Align table columns on decimal point
\usepackage{bm}% bold math
\begin{document}

\preprint{APS/123-QED}

\title{Mesoscopic self-collimation and slow light in all-positive index layered photonic crystals}

\author{Julien Arlandis$^1$, Emmanuel Centeno$^1$, R\'emi Poll\`es$^1$,  Antoine Moreau$^1$, Julien Campos$^{2,3}$, Olivier Gauthier-Lafaye$^{2,3}$ and Antoine Monmayrant$^{2,3}$}
\affiliation{$^1$LASMEA UMR-CNRS 6602 Universit\'e Clermont-Ferrand II, 
Campus des C\'ezeaux, 24, Avenue des Landais, 63177 Aubi\`ere Cedex, France \\ $^2$CNRS ; LAAS ; 7 avenue du colonel Roche, F-31077 Toulouse Cedex 4, France \\
$^3$Universit\'e de Toulouse ; UPS, INSA, INP, ISAE ; UT1, UTM, LAAS ; F-31077 Toulouse, Cedex 4, France 
}

\date{\today}

\begin{abstract}
We demonstrate a mesoscopic self-collimation effect in photonic crystal superlattices consisting of a periodic set of all-positive index 2D photonic crystal and homogeneous layers. We develop an electromagnetic theory showing that diffraction-free beams are observed when the curvature of the optical dispersion relation is properly compensated for. 
This approach allows to combine slow light regime together with self-collimation in photonic crystal superlattices presenting an extremely low filling ratio in air. 

 \end{abstract}
\maketitle

Electrodynamic theory in negative-index materials, originally introduced by V.G. Veselago in 1960 \cite{Veselago:1968p3346}, has known impressive progress since the advent of metamaterials and photonic crystals \cite{Lourtioz:2008p5784}. Both approaches enable an invaluable light control which originates for the former from the realization of effective electric and magnetic permittivities and for the latter from versatile photonic dispersion relations. Many concepts developed for metamaterials in linear and nonlinear regimes have been transposed to PhCs devices since they enable the design of devices working in the visible range of frequency \cite{Vynck:2009p3011, Fabre:2008p2219, Centeno:2007p170}. The frontier between PhCs and metamaterials has become fuzzier with recent emergence of Bragg mirrors including negative-index materials. These composite one-dimensional PhCs, alternating positive- and negative-index layers, have in particular shown an intriguing photonic band gap when around a particular frequency, the average refractive index over one lattice period is null \cite{Li:2003p3780, Yuan:2006p4417, Bria:2004p4795, Shadrivov:2009p4794, Nefedov:2002p4092, Panoiu:2006p3781}. This forbidden range of frequency named zero-$\bar{n}$ gap is for example insensitive to the thickness of the lattice period or disorder \cite{Kocaman:2009p3775} and electromagnetic signals can be tunneled without any phase delay because of a phase compensation mechanism \cite{Kocaman:2011p5859}. Resonant modes lying in the zero-$\bar{n}$ gap can furthermore appear without breaking the lattice symmetry when a Fabry-Perot optical condition is satisfied \cite{Li:2003p3780}. Surprisingly, it has been demonstrated that zero-average index metamaterials supporting these resonant modes can either propel self-collimated beams or focalize optical signals \cite{Polles:2011p5453}. These properties have been explained by introducing a harmonic average index parameter that takes null or negative values while the average index is kept to zero. However, this approach may be difficult to implement experimentally since an accurate control of the metamaterial optical dispersion is required \cite{Polles:2011p5453,Silvestre:2009p4416}. This drawback is circumvent in an alternative device proposed by V. Mocella and co-authors which consists of a periodic stack of 2D photonic crystals (PhCs) and air-layers of equal thicknesses \cite{Mocella:2009p3778}. The PhC slabs were designed to behave as flat lenses of -1 effective index originating from an isotropic dispersion relation obtained with the second photonic band. In that case, self-collimation can be interpreted as successive focalizations of the beam by a lensing effect. Experimental data also reveal that sub-wavelength beams can efficiently be relayed over millimeter-scale distance thanks to a smart optimization of the PhC interfaces. This quasi-perfect impedance matching prevents however to open a large zero-$\bar{n}$ gap so that self-collimation was observed close to the band gap edge. 

Nevertheless, to date, this beam shaping mechanisms is restricted to photonic devices realizing a quasi-zero-average index condition by the use of negative-index materials and this is achieved in structures with high filling factor in air (of 76$\%$) \cite{Mocella:2009p3778}. In this Letter, we demonstrate that mesoscopic self-collimation also arises in photonic crystal superlattices of all-positive index materials and when the zero-average index condition does not hold. Diffraction-free beams are demonstrated when the curvatures of the photonic dispersion curves of the successive media are offset. We propose an electromagnetic theory that enables the design of Bragg media of extremely low filling factor in air or presenting slow self-collimated light. Let us start with an infinite 2D PhC consisting of a square lattice of air holes (lattice constant $a$, radius $r/a=0.2$) etched in a dielectric medium of optical index 2.9. This 2D PhC has an air filling factor of around $12\%$. The first photonic band, computed with the plane wave expansion method \cite{Johnson:2001p5732}, shows flat iso-frequency curves (IFCs) of quasi null curvature around the reduce frequency $a/\lambda=0.235$, Fig.1.  At this frequency, all the waves constituting a beam propel with parallel group velocities ($\bf v_g=grad_k{\omega(\bf k)}$) pointing in the $\Gamma M$ direction revealing the self-collimation effect \cite{Kosaka:1999p3263}. At lower frequencies (red area in Fig.1), the IFCs have a positive curvature leading to a diverging beam, as in an homogeneous medium. Conversely, for higher frequencies, the negative curvature results in beam focusing. Note that in this regime, negative refraction is not linked to a negative phase index related to opposite phase and group velocities since here $\bf k. \bf v_g>0$ for all Bloch waves \cite{Luo:2002p146}. These remarks show the crucial role of the local curvature of the IFCs for beam shaping operations \cite{Hamam:2009p5787}. 

\begin{figure}
\centerline{\includegraphics[width=7.5cm]{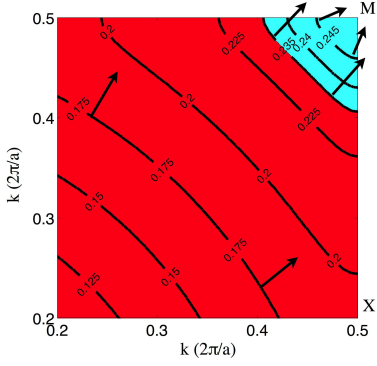}}
\caption{\label{fig:fig1} (color). Iso-frequency curves of the first band for TM polarization. The arrows indicate the direction of $\bf v_g$ and the red and blue areas correspond respectively to IFCs of positive and negative curvatures.}
\end{figure}

Suppose now a PhC superlattice consisting of a periodic set of the previous 2D PhC sized in slabs of thickness $d_1$ and of homogeneous layers of thickness $d_2$ and refraction index $n_2=2.9$. The $\Gamma M$ direction of the 2D PhC is chosen parallel to the  direction of the PhC superlattice stack. Mesoscopic self-collimation should appear when the positive and negative curvatures of the dispersion relations of the homogeneous and the PhC layers are exactly balanced. At the scale of the Bragg crystal macro-period, $D=d_1+d_2$,  the resulting average curvature should vanishes provided that appropriate layers thicknesses are chosen. This concept is developed in the framework of a beam propagation model. Consider a Gaussian beam of initial waist $W_0$ launched in a PhC superlattice presenting a ratio in PhC layers $\alpha=d_2/d_1$. After a unit cell distance $D$, the beam is derived by the inverse Fourier transform of the product of the initial Gaussian beam envelop $U^i(k_x)=W_0/(2\sqrt{\pi})\exp(-(k_x W_0/2)^2)$ by the phase propagators of each media: 
\begin{equation}
U(x,D)=TF^{-1}\left\{U^i(k_x)P_{PhC}(k_x,d_1)P_{hom}(k_x,d_2)\right\}
\label{eq:U}
\end{equation}
where the general expression of the phase propagator in a medium of dispersion relation $k_y(k_x)$ is $P(k_x,y)=\exp[ik_y(k_x)y]$. In high symmetry directions (such as in $\Gamma M$ direction), the parity of the IFCs imposes a null first derivative ${\partial k_y}/{\partial k_x}$. Consequently, the Taylor expansion of the wavevector $k_y$ written as: 
\begin{equation}
k_y(k_x)=k_y(k_{x0})-(k_x-k_{x0})^2 \frac{1}{2\ddot{k}}
\label{eq:dl}
\end{equation}
The right hand term is linked to the local curvature of the IFC by $\left.\frac{1}{\ddot{k}}=-\frac{\partial^2k_y}{\partial{k_x}^2}\right|_{k_{x0}}$. For the sake of simplicity the reciprocal basis is chosen as $k_{x0}=0$ in $\Gamma M$ direction and we define $k_{y0}=k_y(k_{x0})$. In that case, for the homogeneous slab, the curvature is related to the refraction index by $ 1/{\ddot{k}_{hom}}=1/(k_0 n_2) $  ($k_0$ being the wavenumber in vacuum). Similarly, let introduce for 2D PhC structures the dimensionless parameter $n_c$ named  curvature index which is proportional to the curvature radius of the IFCs: $n_c=\ddot{k}_{PhC}/k_0$. As depicted in Fig.2a, $1/n_c$ vanishes at the self-collimation frequency $a/\lambda=0.235$ and takes positive or negative values at lower or higher frequencies respectively. The expansion of Eq.(\ref{eq:dl}) (assimilated to the paraxial assumption) allows one to express the phase propagator over one unit cell: 
\begin{equation}
P_{cell}(k_x,D)=\exp \left[iD \left(\left\langle k_{y0}\right\rangle-\frac{k_x^2}{2k_0}\left\langle\frac{1}{n_{c}}\right\rangle \right) \right]
\label{eq:P}
\end{equation}
which is simply the product $P_{hom}P_{PhC}$. This phase propagator involves the average phase $\left\langle k_{y0}\right\rangle$ and the harmonic average curvature index given by $\left\langle\frac{1}{n_c}\right\rangle=\frac{1}{D}\int_0^D\frac{dl}{n_c(l)}$. Using Eq.(\ref{eq:U}) and Eq.(\ref{eq:P}) one can find an analytical expression of the beam inside the Bragg media. The waist of the Gaussian beam after one macro-period $D$ depends in particular on the average curvature:
\begin{equation}
W(D)=W_0\sqrt{1+\left (\frac {\theta_0D}{W_0} \right)^2 \left\langle\frac{1}{n_c}\right\rangle^2}
\label{eq:W}
\end{equation}
where $\theta_0=\lambda/(\pi W_0)$ is the divergence angle \cite{Born:2000p5388}. This result demonstrates that the initial beam waist is recovered after propagating through one Bragg period, $W(D)=W_0$, when the average curvature is zero:
 \begin{equation}
 \frac{d_1}{n_c}+\frac{d_2}{n_2}=0
 \label{eq:courb}
\end{equation}
Note that here mesoscopic self-collimation is based on the exact balance of the local IFC curvatures of each layers and does not require to open a zero-$\bar{n}$ gap that implies $\left\langle k_{y0}\right\rangle=0$ \cite{Li:2003p3780, Polles:2011p5453}.
\begin{figure}
\centerline{\includegraphics[width=7cm]{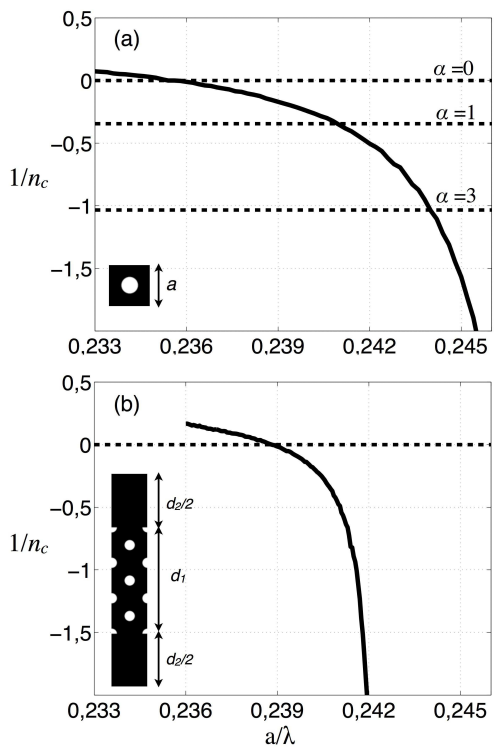}}
\caption{\label{fig:fig2} Curvature function $1/n_c$ computed in $\Gamma M$ direction versus the reduce frequency. (a) For the infinite 2D PhC  described in Fig.1 and of unit cell depicted in the inset. The intersection points between $1/n_c$ and the  straight dashed-lines for $\alpha=0$, $1$ and $3$ correspond respectively to the self-collimation frequencies for the unbounded 2D PhC and for PhC superlattices of two filling ratios in PhC layers. (b) Graph of $1/n_c$ computed for the supercell corresponding to a PhC superlattice with $\alpha=1$ and $d_1=3a\sqrt2$ shown in the inset. Mesoscopic self-collimation arises for $a/\lambda=0.239$  where $1/n_c$ vanishes.}
\end{figure}
The beam propagation model in PhC superlattices provides a simple method to determine the self-collimation frequency from the 2D infinite PhC dispersion properties. The working frequency is indeed fixed by the intersection point of the local curvature $1/n_c$  with the straight line $- \alpha/n_2$ deduced from Eq.(\ref{eq:courb}), Fig. 2. For ratios in PhCs layers $\alpha=1$ and $\alpha=3$, mesoscopic self-collimation should respectively be obtained at the reduced frequencies $a/\lambda=0.241$ and $a/\lambda=0.244$, Fig.2a. The operating frequency is indeed blue shifted when increasing $\alpha$ because the PhC layers must focus more strongly to cancel diffraction in thick homogeneous slabs. It is also worth noting that these frequencies are invariant with respect to the macro-period $D$ provided that Eq.(\ref{eq:courb}) is satisfied. This invariant scaling property has been checked by calculating the photonic band diagrams of PhC superlattices with the plane wave expansion method. Flat IFCs are found for a constant $\alpha$ and for several $d_1$ varying from $a\sqrt2$ to $4a\sqrt2$ (by step of $a\sqrt2$). For example, it is seen in Fig.2b that for $d_1=3a\sqrt2$ and $\alpha=1$, the curve $1/n_c$ vanishes at $a/\lambda=0.239$ in agreement with the beam propagation model. The weak difference (smaller than $1\%$) between this operating frequency and the prediction of the propagation beam model shows that our model is thus accurate enough to guide designing and fabrication of self-collimated PhC structures. To corroborate these results, the propagation of an incident beam has been performed with 2D FDTD simulations through PhC superlattices presenting two filling ratios $\alpha$ \cite{Farjadpour:2006p5783}. We found a self-collimated beam over a length of $400a$ at the reduce frequencies $0.239$ and $0.243$ for $\alpha=1$ and $3$ respectively, Fig.3. Theses results demonstrate that mesoscopic self-collimation is achieved in layered structures of extremely low filling factor in air. The air filling factor given by $f=\pi (r/a)^2/(1+\alpha)$ in the superlattices of Fig.3a and 3b is respectively divided by 2 and 4 versus to the initial unbounded 2D PhC. Compared to the bulk semiconductor case where the beam waist attains 3 times $W_0$ after the same distance $400a$ (Fig.3c), the beam keeps the same initial waist for the PhC superlattice presenting an air filling ratio of only $3\%$ (Fig.3b). This value could furthermore be decreased by considering 2D PhC layers of lower curvature index. The selectivity of PhC superlattices can be evaluated by computing the range of frequency supporting the self-collimation effect in the same way than in \cite{Hamam:2009p5787}. Here, in the framework of the beam propagation model, it is related to the derivative of the waist versus the reduce frequency which is derived for a large propagation distance $D$ from Eq. (\ref{eq:W}) : $\partial W/\partial (a/\lambda)=\theta_0 d_1 \partial(1/n_c)/\partial (a/\lambda)$. The selectivity is then driven by the slope of  the curvature $1/n_c$ of the infinite PhC and it increases with higher values of $a/\lambda$, see Fig. 2a. For example, the self-collimation bandwidth is two times smaller for $\alpha=1$ compared to the infinite PhC case. Conversely to the approach proposed in \cite{Hamam:2009p5787}, the self-collimation bandwidth is reduced for PhC superlattices working with the first photonic band. This conclusion may however be reconsidered for others photonic dispersion relations.  
\begin{figure}
\centerline{\includegraphics[width=7cm]{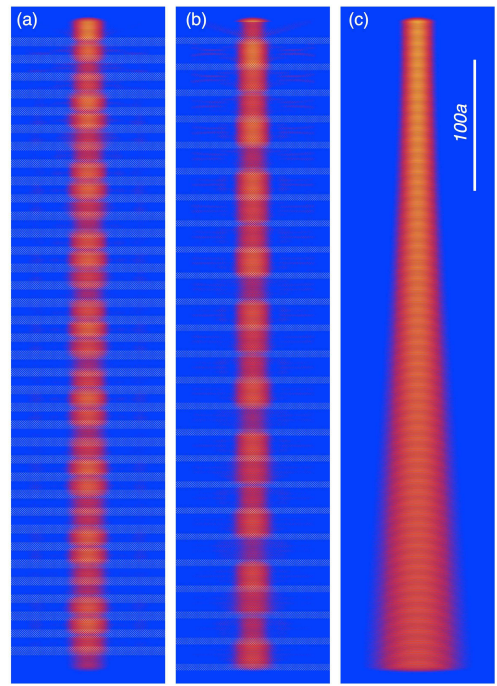}}
\caption{\label{fig:fig3} (color) Map of the magnetic field modulus for a Gaussian beam of waist $W_0=8a$ (a) At the reduce frequency $a/\lambda=0.239$ for a PhC superlattice for $d_1=4a\sqrt2$ and $\alpha=1$. (b) At the reduce frequency $0.243$ for $d_1=3a\sqrt2$ and $\alpha=3$. (c) The same Gaussian beam propagating in a homogeneous medium of optical index $n_2=2.9$ and for $a/\lambda=0.239$.}
\end{figure}

\begin{figure}
\centerline{\includegraphics[width=7.5cm]{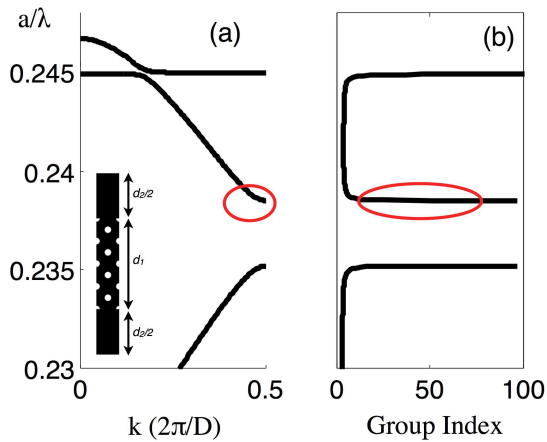}}
\caption{\label{fig:fig4} (color). (a) Band diagram in the $\Gamma M$ direction of the PhC superlattice for $\alpha=1$ and $d_1=4a\sqrt2$ depicted in the inset. The red circles indicate the location of the self-collimation frequency $0.239$. (b) Group index versus the reduce frequency.}
\end{figure}

Besides of these linear properties, these PhC superlattices present also attractive properties for nonlinear applications such as laser emission or frequency conversion. Nonlinear optical processes developed for Bragg mirrors can be revisited owing to the original optical properties of PhC superlattices. First, the latter enables to guide self-collimated beam in devices presenting a strong amount of non-etched semiconductor material. Indeed, as compared to structures reported to date \cite{Mocella:2009p3778}, the air filling factor is reduced by a factor 25 resulting in a potentially $97\%$ of optically active or nonlinear semiconductor. Moreover, as seen in Fig.3b, the PhC superlattice of low ratio in PhC layers behaves like a set of coupled cavities where a strong electromagnetic field localizes. Second, light-matter interaction is also enhanced when light is slowed down at frequencies close to a photonic band gap (PBG) edge \cite{Baba:2008p3148}. In lasers, the optical gain increases linearly with the group index, $n_g=c/v_g$, and the efficiencies of second or third harmonic generation nonlinear processes are enhanced as $n_g^2$.  Unfortunately, self-collimation in 2D or 3D PhCs arises far away from the PBG edge preventing thus to work in a slow light regime. In our case, the infinite square lattice of air holes presents a complete PBG between the first and the second bands spanning the frequency range 0.247-0251. Self-collimation arising at the reduce frequency 0.235 is associated to Bloch modes of around 2.96 group index which is almost the refraction index of the host material. To increase this low group index one can play with the additional degree of freedoms provided by PhC superlattices. Since these layered media are basically Bragg mirror structures, we can open an additional PBG at the vicinity of the mesoscopic self-collimation frequency by an appropriate choice of the macro-period $D$ and the parameter $\alpha$. For example, the band diagram computed in the $\Gamma M$ direction, for $\alpha=1$ and $d_1=4a\sqrt2$ corresponding the structure of Fig.3a, reveals that a new PBG centered at $a/\lambda=0.237$ and originating from Bragg scattering opens, Fig.4a. With these parameters, the self-collimation frequency, $a/\lambda=0.239$, appears now at the band gap edge located at the reduce frequency $0.2385$. A high group index of around 50 is observed for a PhC superlattice of $6\%$ air filling factor that combines both slow light and the self-collimation effects, Fig.4b.

In summary, a theory of photonic dispersion curvature compensation has been presented to predict self-collimation of light in PhC superlattices. The beam propagation model is shown to be relevant for the conception of PhC superlattices and gives an insight on the optical mechanism behind the mesoscopic self-collimation effect. This theory reveals that negative-index materials are not necessary to generate diffraction-free beams leading thus to a general approach for the design of Bragg mirrors based on 2D photonic crystals properties. Mesoscopic self-collimation is demonstrated in PhC superlattices presenting $97\%$ of non-etched semiconductor.  Slow self-collimated light is in addition showed in appropriately designed PhC superlattices that work at a frequency located at the edge of an additional photonic band gap. To our opinion, these novel properties pave attractive avenues for the design of active or nonlinear devices presenting beam shaping functionalities. 
~\\

This work is supported by the Agence Nationale pour la Recherche of France ($NT09\_517025$). The authors thank the Centre Informatique National de l'Enseignement Sup\'erieur for its computing support.

\end{document}